# Experimental study of magneto-superconductor $RuSr_2Eu_{1.5}Ce_{0.5}Cu_2O_{10-\delta}$ :
## Effect of Mo doping on magnetic behavior and $T_c$ variation

V.P.S. Awana, R. Lal and H. Kishan

National Physical Laboratory, K.S. Kirishnan Marg, New Delhi-110012, India

A.V. Narlikar[$], M. Peurla and R. Laiho

Wihuri  Physical Laboratory, University of Turku, FIN - 20014, TURKU, Finland.

[$]Also at UGC-DAE Consortium for Scientific Research, University Campus, Khandwa Road, Indore-452017, MP, India.

Mo doped ruthenocuprates $Ru_{1-x}Mo_xSr_2Eu_{1.5}Ce_{0.5}Cu_2O_{10-\delta}$ are synthesized for x = 0.0, 0.2, 0.4, 0.6, 0.8 and 1.0, and their magnetic and superconducting properties are studied. It has been found that the magnetic transition temperature $T_{ZFC}^{peak}$ , which corresponds to the appearance of weak ferromagnetic effect, decreases from its value of 75 K for x = 0.0 to 22 K, 25 K and 18 K, respectively for the x = 0.2, 0.4 and 0.6 samples. Another finding is that the magnetic susceptibility reduces at $T_{ZFC}^{peak}$ by a factor of about 6, 85 and 413 for x = 0.2, 0.4, and 0.6 respectively. The samples of x = 0.8 and 1.0 are found to have no magnetic or superconducting effects. The values of the superconducting transition temperature are obtained from the resistivity versus temperature data. An important result is that $T_c$ increases by 4.5 K and 7.0 K for x = 0.2 and 0.4 respectively, and then decreases by 17 K for x = 0.6.  The observed variation of $T_c$ with x has been explained in terms of a theory which combines the effects of weakening magnetic behavior and reducing carrier concentration in a phenomenological manner. The resulting theory is found to provide a good agreement with the observed value of $T_c$.

Key words: Ruthenocuprates, Superconductivity and Magnetism, and Magneto transport

## I.        INTRODUCTION

Coexistence of superconductivity and magnetism (in particular the ferromagnetism) within a single (thermodynamical) phase has been a point of discussion over decades [1, 2]. Early evidence of these two coexisting phenomenon were realized in various f-electron compounds viz. $ErRh_4B_4$ [3,4]. Recently co-existence of superconductivity and magnetism has also been observed in $UGe_2$ [5], $ZrZn_2$ [6], and in ruthenocuprates viz. $RuSr_2GdCu_2O_8$ (Ru-1212) and $RuSr_2(Gd,Sm,Eu)_{1.6}Ce_{0.4}Cu_2O_{10-\delta}$ (Ru-1222) [7,8,9]. In the ruthenocuprates, magnetism originates basically from the $RuO_2$ sheets, and superconductivity arises from the $CuO_2$ layers. Experiments like ESR, NMR and magneto-optics show genuine coexistence of superconductivity and magnetism within the same phase of these compounds [8, 10, 11]. Some workers have also considered the possibility of phase separation, i.e., isolated superconducting and magnetic regions within the same compound [12-15].

More information on the interacting magnetism and superconductivity in ruthenocuprates can be gained from such ruthenocuprates where Ru is replaced partially by other ions of magnetic (Co, Fe) or non-magnetic (Nb, Mo) nature. A Study of the change in the magnetic properties of such a composite system is expected to provide useful information. In this regard there are already some reports on $Ru_{1-x}M_x$:1222 [16-18], where M =Fe, Co, Nb and Mo. In this paper we consider substitution of Ru by Mo, and measure electric and magnetic behavior for various contents of Mo. It may be noted that both Ru-1222 and Mo-1222 are isostructural [16-19], so we do not expect complications from the lattice effects in the doped system. We study electrical and magnetic properties of $Ru_{1-x}Mo_xSr_2Eu_{1.5}Ce_{0.5}Cu_2O_{10-\delta}$ ($1.0 \geq x \geq 0.0$) magnetosuperconductor. We expect these properties to be affected by Mo due to its two main roles. The first is the decrease of carrier density by the Mo substitution because Mo has valency +6, while Ru has valency +5. The second role of Mo lies in diluting the magnetic effect of Ru. This is expected to result in the increased superconducting transition temperature $T_c$, as the low $T_c$ of ruthenocuprates is due to partial suppression of superconductivity by the magnetism caused by Ru ions.

## II.  EXPERIMENTAL DETAILS

Samples of $Ru_{1-x}Mo_xSr_2Eu_{1.5}Ce_{0.5}Cu_2O_{10-\delta}$ ($1.0 \geq x \geq 0.0$) series were synthesized through a solid-state reaction route from stoichiometric amounts of $RuO_2$, $SrO_2$, $Eu_2O_3$, $CeO_2$, CuO and $Mo_2O_3$. Calcinations were carried out on mixed powders at 1000 ℃, 1020 ℃ and 1040 ℃ for 24 hours at each temperature with intermediate grindings. The pressed bar-shaped pellets were annealed in flow of oxygen at 1075 ℃ for 40 hours and subsequently cooled slowly over a span of another 20 hours down to room temperature. The same pellets were further annealed in flow of $O_2$ at one atmospheric pressure at $400^0C$ for 24 hours and slowly cooled to room temperature in same environment over a span of 6 hours. X-ray diffraction (XRD) patterns were collected at room temperature with Cu$K_\alpha$ radiation. Resistivity measurements were carried out by conventional four-probe method on a close-cycle-referigerator (CCR) down to 12K. Magnetization measurements were performed on a SQUID magnetometer (Cryogenic Ltd. model S600).

## III.  RESULTS AND DISCUSSION

### A. x-ray diffraction

X-ray diffraction (XRD) patterns for $Ru_{1-x}Mo_xSr_2Eu_{1.5}Ce_{0.5}Cu_2O_{10-\delta}$ ($1.0 \geq x \geq 0.0$) series of samples are shown in Fig. 1. It is evident from this figure that all samples crystallize in single phase with tetragonal structure (space group *I4/mmm*). Respective indices are shown in the figure. As far as pristine x = 0.0 is concerned, very small amounts of $SrRuO_3$ and/or $GdSr_2RuO_6$ are seen close to main intensity peaks which had earlier been noted by other workers also [7, 11, 12, 16]. Our currently studied samples are in fact, far better in terms of their phase purity as compared to those reported earlier by various authors. As far as the position (2θ) of main peaks are concerned, small decrease in them is seen due to lower ionic size of $Mo^{6+}$ than of the $Ru^{5+}$ ion. This is in agreement with an earlier report on the same system [16, 19]. The pristine compound (x = 0) has its lattice parameters $a = b = 3.835$ (2) Å and $c = 28.493$ (6) Å. On the other hand lattice parameters of the fully Mo substituted (x = 1) compound are $a = b = 3.843$ (5) Å and $c = 28.4653$ (4) Å. For Mo concentrations higher than x = 0.6, some extra peaks (marked with *) of small intensity, along with $SrMoO_4$ and $CuO$ are also seen. Despite the presence of small intensity un-reacted peaks, the majority of $Ru_{1-x}Mo_xSr_2Eu_{1.5}Ce_{0.5}Cu_2O_{10-\delta}$ compounds are mainly single phase.

**B. Magnetic behavior**

In Fig. 2 we show the magnetization $M$ of ruthenocuprate $Ru_{1-x}Mo_xSr_2Eu_{1.5}Ce_{0.5}Cu_2O_{10-\delta}$ at different values of the applied magnetic field $H$. In fact, we have applied magnetic fields up to 6 T. The results are shown for the superconducting samples (x = 0, 0.4 and 0.6) and for the nonsuperconducting sample (x = 0.8) at the temperature value of $T = 5$ K. The observed behavior of $M$ may be expressed as a sum of two contributions:

$$M = \chi H + \sigma_s (H) \qquad (1)$$

The first term on the right-hand-side of this equation is linear in the magnetic field $H$ with $\chi$ as dc susceptibility. This contribution arises from the antiferromagnetic (AFM) or spin glass behavior of the Ru spins. The second term in the right-hand-side of Eq.(1), $\sigma_s (H)$, represents the ferromagnetic (FM) component of the Ru spins. From Fig. 2 it is clear that the x = 0 sample is characterized mainly by the FM component. In fact, if we extrapolate the $M$ vs. $H$ curve down to $H = 0$ T, it turns out that $M$ takes the value 3.9 emu/g. This means that the x = 0 ruthenocuprate has an essential presence of the FM effect in it. On the other hand the x = 0.8 sample is completely linear in $H$ in the complete range of the considered values of $H$. This means (cf. Eq.1) that the FM component $\sigma_s (H) = 0$ for this sample. When we move from the x = 0.0 sample to the x = 0.8 sample via the x = 0.4 and 0.6 samples, we observe from Fig. 2 that the FM component $\sigma_s (H)$ gets weakened with increasing x, while the linear component $\chi H$ is enhanced. On this basis it may be said that with the increasing substitution of the Ru atoms by the Mo atoms the FM effect is diluted gradually so that at the 80% substitution by Mo atoms the FM effect is completely lost.

In Fig. 3 we present the plots of the temperature-dependence of the dc susceptibility $\chi$ for x = 0.0, 0.2, 0.4 and 0.6. In the inset of this figure we show enlarged $\chi$ versus $T$ plot for x=0.2 to show , in particular, branching of the field-cooled (FC) and zero-field-cooled (ZFC) curves near 135 K. Formally, the shapes of the field-cooled and zero-field-cooled curves of Fig. 3 for different values of x are similar to that for x = 0 [20]. The inset makes it clear for x=0.2. In order to compare the $T$-dependence of $\chi$ for various samples we present, in table I, values of $T_{mag}$, $\chi_{ZFC}^{peak}$ and $T_{ZFC}^{peak}$. Here $T_{mag}$ is the temperature where AFM effect starts to occur when temperature is lowered from the higher side. $T_{mag}$ is identified from the separation of the FC and ZFC branches. $\chi_{ZFC}^{peak}$ denotes the maximum value of the susceptibility for the ZFC case. This value of $\chi$ signifies the appearance of a weak FM/spin-glass(SG) state [21]. The weak FM state is believed to originate from the canting of the Ru moments, which, in turn, results probably from an antisymmetric exchange coupling of the Dzyaloshinsky-Moriya type [22]. The temperature at which $\chi$ has a peak in the ZFC branch is denoted by $T_{ZFC}^{peak}$.

From table I we see that the temperature $T_{mag}$ decreases slowly with increasing x. On the other hand $T_{ZFC}^{peak}$ shows a sharp and complicated (non monotonic) behavior with increasing Mo content. First of all, $T_{ZFC}^{peak}$ drops sharply, by 72%, for the x = 0.2 sample. Then for the x = 0.4 sample it increases slightly (by 3 K) with respect to that for the x = 0.2 sample. For the x = 0.6 sample, value of $T_{ZFC}^{peak}$ falls again. As far as the variation of $\chi_{ZFC}^{peak}$ for different x, we see from Table I that $\chi_{ZFC}^{peak}$ is reduced by a factor of about 6, 85 and 413 for the x = 0.2, 0.4, and 0.6 samples respectively. Since the x = 0 sample is already a weak FM material [21], this strong reduction of the values of $\chi_{ZFC}^{peak}$ with x implies that the FM effect disappears rapidly with increasing Mo content.

### C. Resistivity and Superconductivity

Resistivity ($\rho$) versus temperature plots for the Mo-doped samples of $RuSr_2Eu_{1.5}Ce_{0.5}Cu_2O_{10-\delta}$ are shown in Fig. 4 for temperature values up to 160 K. Only the superconducting samples (x = 0.0, 0.2, 0.4, 0.6) are shown in Fig. 4. The samples with 80% Mo and 100% Mo are not superconducting. It is clear from Fig. 4 that all the samples show upturn of resistivity near the onset transition temperature $T_c^{onset}$. The extent of upturn of $\rho$ increases sharply for the x = 0.2 sample in comparison to that of the x = 0 sample. The x = 0.4 sample has almost the same $T$-dependence of $\rho$ (up to 160 K) as for the x = 0.2 sample. But the extent of the upturn to $\rho$ near $T_c^{onset}$ is once more enhanced sharply for the x = 0.6 sample. The overall behavior to the $\rho$ versus $T$ plots of Fig. 4 is reminiscent of the underdoped cuprate superconductors [23], and is in general, agreement with other reports on the $RuSr_2Eu_{1.5}Ce_{0.5}Cu_2O_{10-\delta}$ superconductor [7, 11, 12, 16].

We estimate the transition temperature $T_c$ for different values of x from Fig. 4 in a way described in Ref. [24]. The values of $T_c$ obtained in this way are shown in Fig. 5 for x = 0.0, 0.2, 0.4 and 0.6. An interesting result from Fig. 5 is that up to x = 0.4 Mo enhances $T_c$ with increasing x. In fact estimated values of $T_c$ are 33 K, 37.5 K, 40 K and 23 K for x = 0.0, 0.2, 0.4 and 0.6, respectively. In order to understand the behavior of $T_c$ with x we consider the roles of Mo in the ruthenocuprate systems. The first thing which we notice is that Mo is nonmagnetic. So, its partial entry at the position of the Ru sites will tend to weaken the magnetic effect of the Ru moments. As has been discussed above (Fig. 2 and table I) the magnetic effect due to Ru moments is completely destroyed for and above x = 0.8. In order to take account of this role of the Mo ions, we treat the effect of the weakening of the magnetic effect of Ru moments within the Abrikosov-Gorkov (AG) theory [25]. Since the original or weakened magnetic effect is essentially due to the presence of the Ru ions, we should consider the concentration of Ru ions formally as the concentration of magnetic impurities in the AG theory. In this sense, the concentration of the magnetic impurities will be given by 1-x. That is to say, the system is effectively pure for 1-x = 0. Let $T_{co}$ denote the temperature of such a pure system. Then, the transition temperature given by the AG theory must satisfy

$$T_{c,AG}(x) = \begin{cases} T_{co} & \text{(for x=1)} \\ \\ T_c(x = 0) = 33\text{K} & \text{(for x =0)} \end{cases} \qquad (2)$$

The AG theory expresses the temperature $T_{c,AG}$ for a given impurity concentration 1-x through the expression

$$\ln\left(\frac{T_{c,AG}}{T_{co}}\right) = \psi\left(\frac{1}{2}\right) - \psi\left(\frac{1}{2} + \frac{A(1-x)}{2T_{c,AG}}\right) \qquad (3)$$

Here $\psi$ denotes the digamma function, and the parameter A, having dimensions of temperature, is given by

$$A = \frac{7 N_F u_2^2 S (S + 1)}{96 \pi k_B} \qquad (4)$$

Here $N_F$ is the density of states at the Fermi level, $u_2$ is the strength of the exchange interaction caused by the interaction of the carrier holes with the Ru moments, $k_B$ is the Boltzmann constant, and S is impurity (Ru) spin.

Eq. (3) is based only on the weakening of the magnetic effect due to the Mo ions. There is another important effect of the Mo ions. In fact, in ruthenocuprates Mo ions will exist in the $Mo^{+6}$ form, as compared to the $Ru^{+5}$ form of the Ru ions. This means that each Mo atom adds one electron in the system, thereby reducing the concentration of the carrier holes. Since the x = 0 ruthenocuprate appears to correspond to the underdoped regime of the cuprate superconductors (see above), and since in the underdoped regime decrease of carrier concentration amounts to suppression of $T_c$ [26], the effect of Mo atoms will tend to decrease $T_c$. Since the AG theory considers fixed carrier concentration, such a decrease of $T_c$ requires modification of Eq. (3). We do not go into such complications. Rather, we assume that the two effects of the variation of $T_c$ - weakening of the FM effect and decrease of the carrier density - are separable from each other so that $T_c$(x) may be written as

$$T_c\text{ (x)} = T_{c,\text{AG}}\text{ (x) } R\text{(x)} \qquad (5)$$

Here the factor $R$(x) describes the suppression of $T_c$ due to the reduction of the carrier density, and so is defined by

$$R\text{(x)} = \begin{cases} 1 & \text{(for x=0)} \\ \text{decreasing} & \text{(for increasing x)} \\ 0 & \text{(for x} \geq 0.8) \end{cases} \qquad (6)$$

For simplicity we consider the following form for R (x):

$$R\text{ (x)} = \theta\ (0.8 - \text{x})\ [(0.8 - \text{x})/0.8]^{\,a} \qquad (7)$$

Here $\theta$ (y) =1 for y > 0, and $\theta$ (y) = 0 for y < 0. The parameter $a$ is a phenomenological parameter. It is not difficult to see that Eq. (7) satisfies the conditions of Eq. (6).

Eqs. (2), (3), (4) and (7) provide the phenomenology of $T_c$ variation in terms of two parameters- $T_{co}$ and $a$. On the basis of calculations from these equations we have found that the resulting values of $T_c$ agree well with the experimental data for $T_{co} = 90$ K and $a =0.55$. This is shown in Fig. 5. In this figure we have also shown the separate effects due to the AG theory and reduction of carrier concentration. The curve labeled $T_{c,\text{AG}}$ shows the effect of pair breaking only, while the curve marked $T_{c,\text{Mo}}$ shows the effect of carrier reduction only ($T_{c,\text{Mo}} = 33.0$ K). It may be noted that $T_{co}$ is slightly less than the transition temperature of $YBa_2Cu_3O_7$. Moreover, the value of $a$ signifies that initially the effect of reduction of carrier concentration is relatively slower than that for the larger values of x. This is on expected lines [26], and so shows the internal consistency of the above interpretation.

## IV.    CONCLUDING REMARKS

We have synthesized samples of ruthenocuprate $RuSr_2Eu_{1.5}Ce_{0.5}Cu_2O_{10-\delta}$ by substituting Ru ions by the Mo ions of concentration x = 0.2, 0.4, 0.6, 0.8 and 1.0. Phase purity and lattice parameters of these samples are discussed on the basis of the XRD. In order to understand the effect of Mo ions on the magnetism of the ruthenocuprate Ru-1222, we have performed measurements of magnetizaton ($M$) for different values of the applied field ($H$). The $H$-dependence of the magnetization (taken at 5 K) shows that Mo suppresses the weak FM effect of the Ru ions monotonically such that this (FM) effect is completely lost by 80% substitution of Ru by Mo. Similar result is obtained from the values of the susceptibility at the temperature $T_{ZFC}{}^{peak}$ where weak FM starts to appear from the higher temperature side. Although the values of $T_{ZFC}{}^{peak}$ is reduced drastically for all of the samples of finite Mo content, the value of $T_{ZFC}{}^{peak}$ is higher for the x = 0.4 sample than that for the x = 0.2 sample. Another feature of the magnetic behavior is that the temperature where AFM effect starts to appear, $T_{mag}$, reduces rather slowly with x. In fact, up to x = 0.6 the value of $T_{mag}$ has reduced from 140 K to 100 K only.

We have also measured resistivity of the considered samples. Only the samples with x= 0.0, 0.2, 0.4 and 0.6 are found superconducting above 12 K. The behavior of resistivity for all the superconducting samples is similar to that of the underdoped cuprate superconductors. From the $\rho$ -T plots we have extracted the values of the transition temperature $T_c$. It has been found that $T_c$ increases with x up to x = 0.4. Thereafter $T_c$ decreases for the x = 0.6 sample. For the x = 0.4 sample, the value of $T_c$ becomes 40 K as against the value of 33 K found for the x=0 sample. We have made an attempt to interpret the observed variation of $T_c$ with x by combining two different roles of the Mo atoms. The first is the weakening of the FM effect of the Ru moments. This effect is expressed through the AG theory. The second role of the Mo atoms is the reduction of the carrier concentration. This feature is modeled empirically, and a good agreement has been achieved with the experimental data.

Table I: Values of $T_{mag}$, $\chi_{ZFC}{}^{peak}$ and $T(\chi_{ZFC}{}^{peak})$ for x=0.0, 0.2,0.4 and 0.6 samples.

| x | $T_{mag}$ (K) | $\chi_{ZFC}{}^{peak}$ (emu/gm) | $T(\chi_{ZFC}{}^{peak})$ (K) |
|---|---|---|---|
| 0.0 | 140 | $1.98 \times 10^{-2}$ | 78 |
| 0.2 | 135 | $3.3 \times 10^{-3}$ | 22 |
| 0.4 | 120 | $2.34 \times 10^{-4}$ | 25 |
| 0.6 | 100 | $4.8 \times 10^{-5}$ | 18 |

**ACKNOWLEDGEMENT**

Authors from the NPL appreciate the interest and advice of Professor Vikram Kumar (Director) in the present work. One of us (AVN) thanks the University of Turku for providing research facilities and for the invitation for the present visit.

**FIGURE CAPTIONS**

Fig. 1.   x-ray diffraction patterns for the $Ru_{1-x}Mo_xSr_2Eu_{1.5}Ce_{0.5}Cu_2O_{10-\delta}$ system.

Fig. 2.   $M$ versus $H$ for the $Ru_{1-x}Mo_xSr_2Eu_{1.5}Ce_{0.5}Cu_2O_{10-\delta}$ system at 5 K for x = 0.0, 0.4, 0.6, and 0.8.

Fig.3    $\chi$ versus $T$ for the $Ru_{0.8}Mo_{0.2}Sr_2Eu_{1.5}Ce_{0.5}Cu_2O_{10-\delta}$ sample for x = 0.0, 0.2, 0.4 and 0.6. The inset shows an enlarged form of the $\chi$ versus $T$ behavior for x = 0.2.

Fig. 4    $\rho$ versus $T$ plots for the $Ru_{1-x}Mo_xSr_2Eu_{1.5}Ce_{0.5}Cu_2O_{10-\delta}$ system for x = 0.0, 0.2, 0.4 and 0.6.

Fig. 5    Transition temperature versus x plots for the $Ru_{1-x}Mo_xSr_2Eu_{1.5}Ce_{0.5}Cu_2O_{10-\delta}$ system. The symbol * shows the experimental points, and the curve marked with $T_c(x)$ shows the final result of the interpretation presented in the text.  The separate effects of pair breaking and reduction of carrier concentration are shown by dashed-dotted line (-.-., $T_{c,AG}$ ) and dashed line (---, $T_{c,Mo}$ ) respectively.

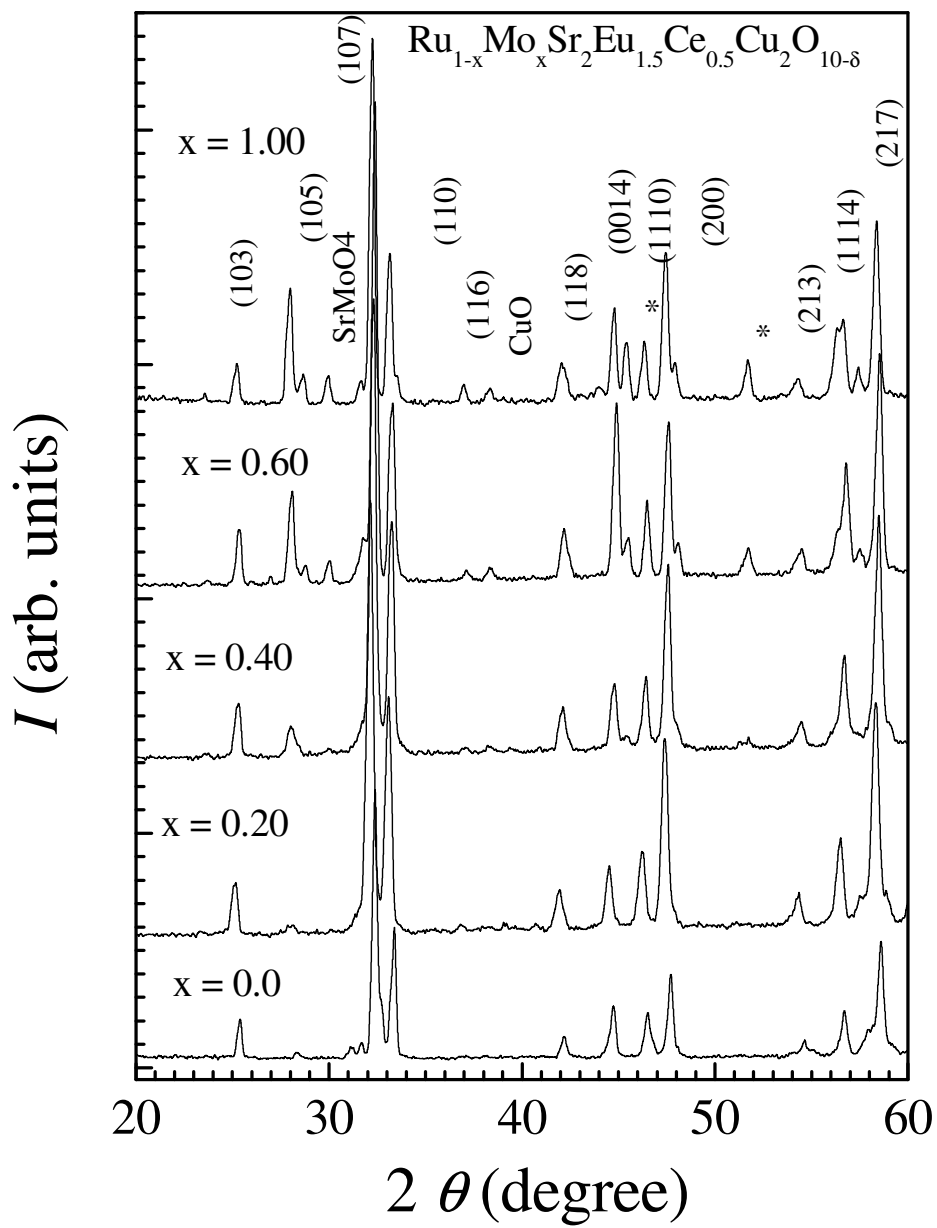

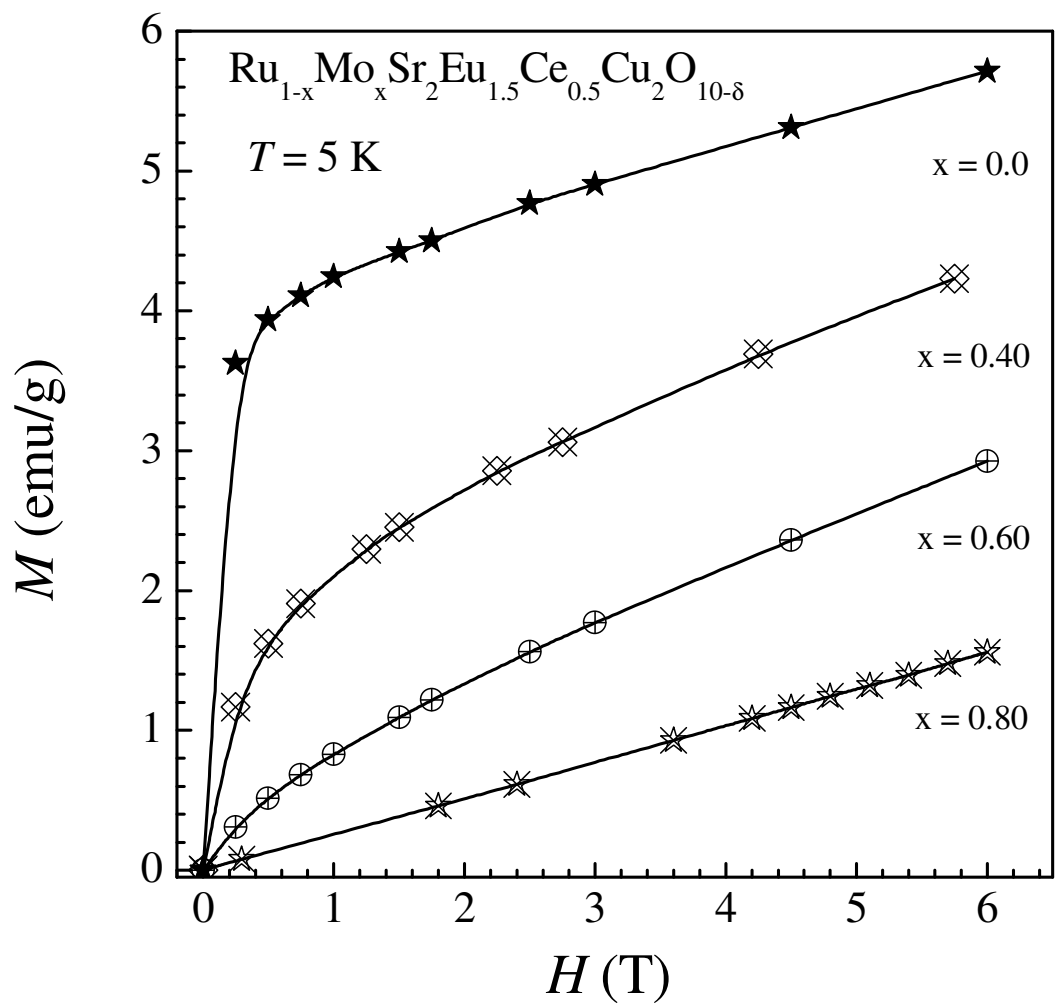

Fig. 3 Awana et al.

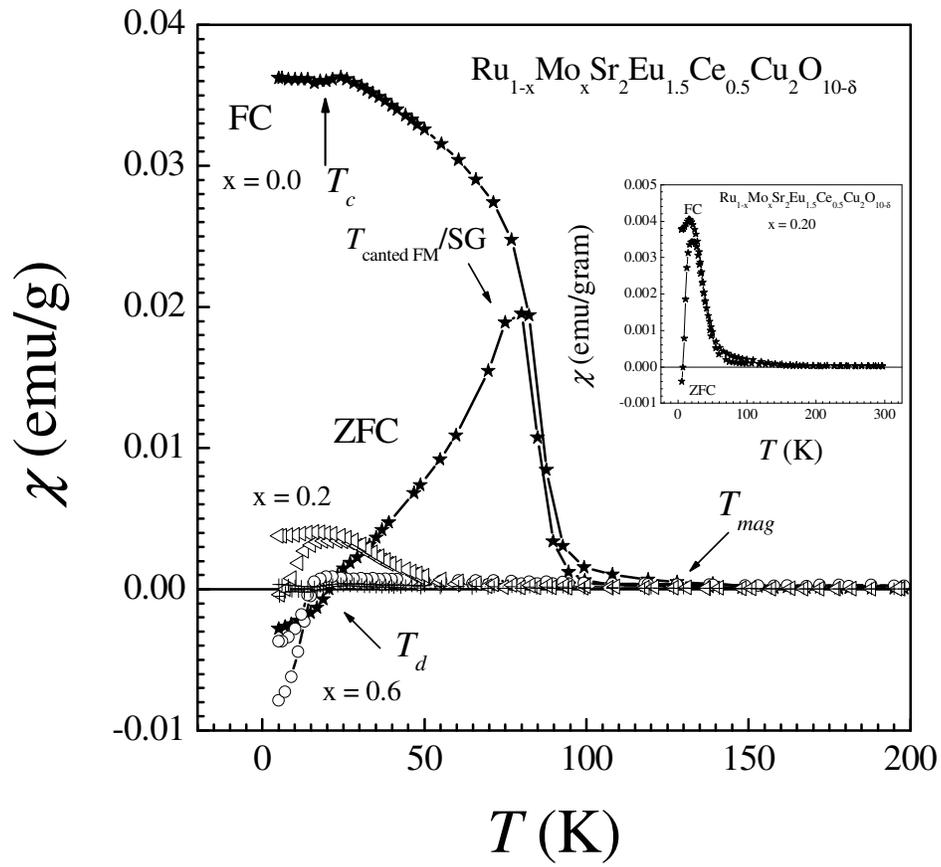

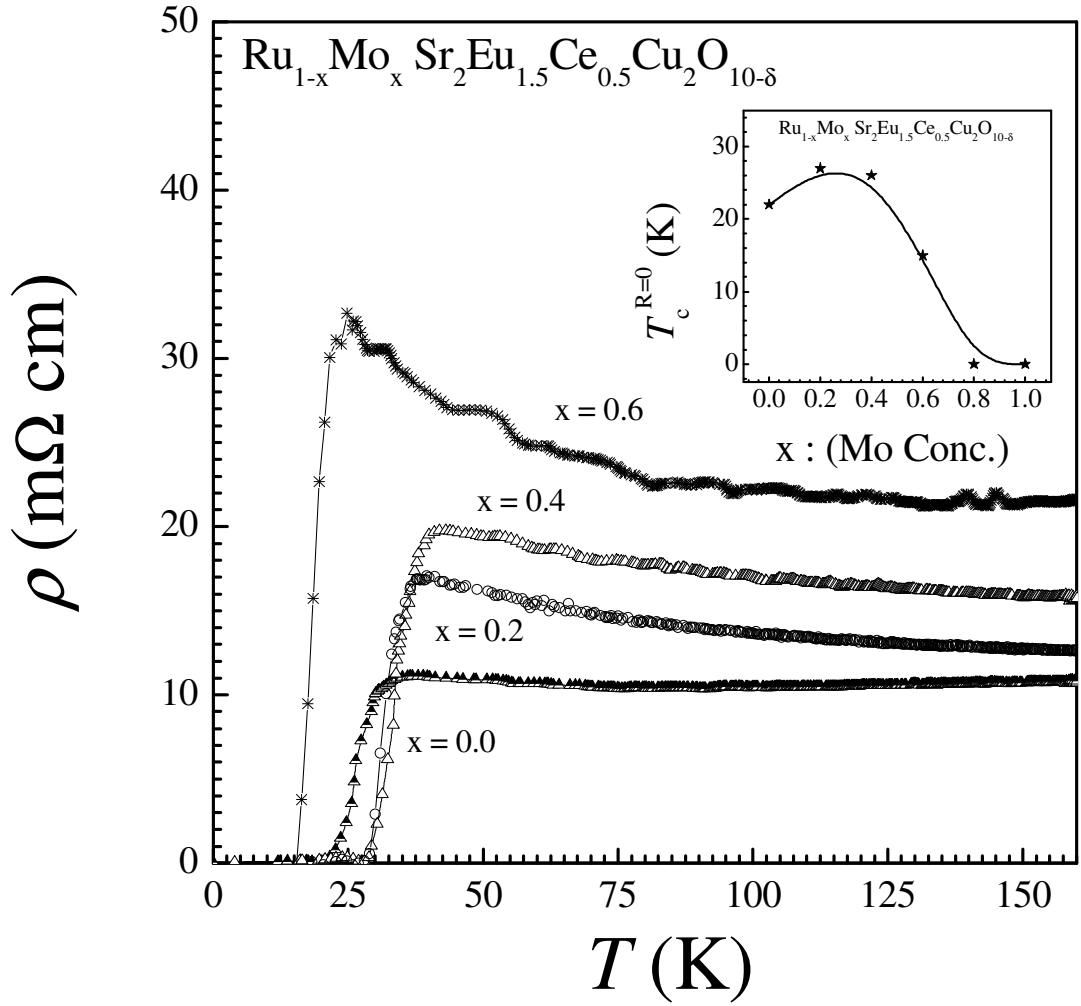

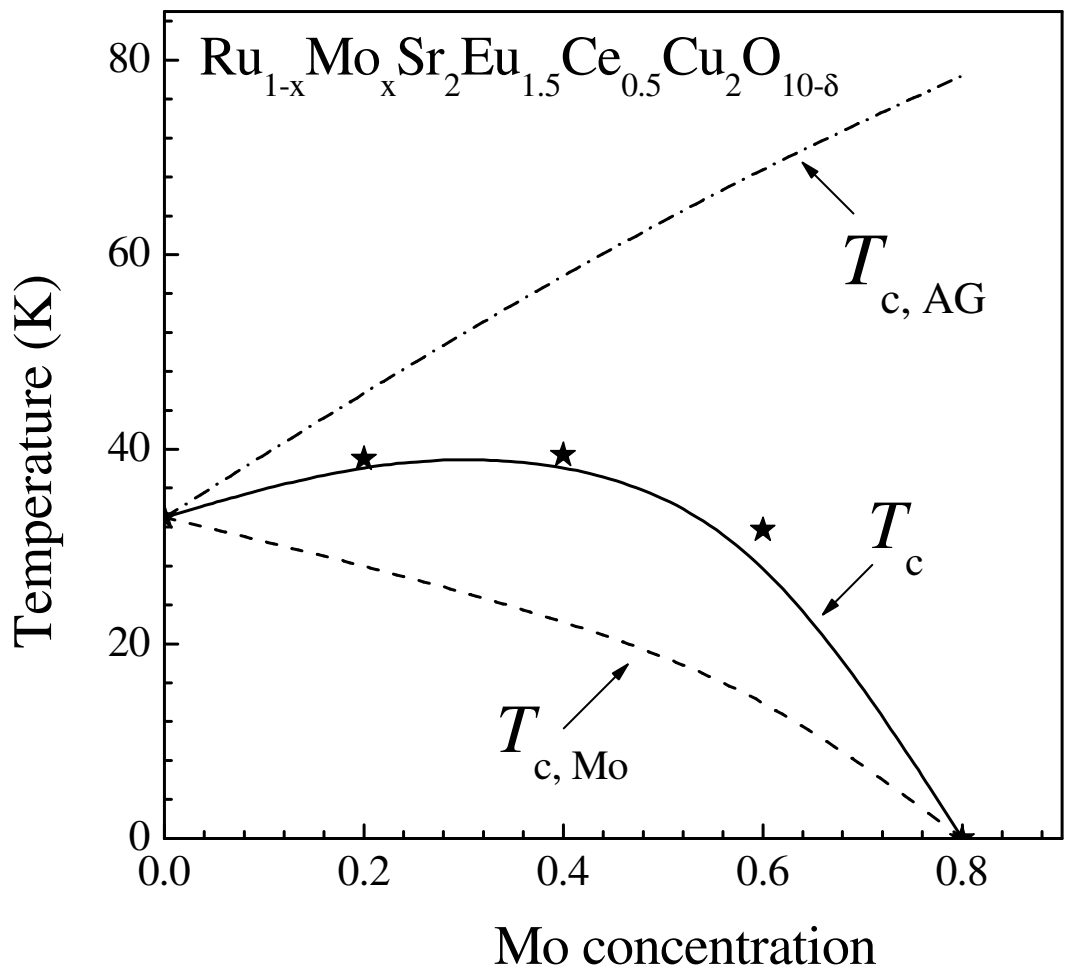

$Ru_{1-x}Mo_xSr_2Eu_{1.5}Ce_{0.5}Cu_2O_{10-\delta}$